\documentclass[%
 aip,
 amsmath,amssymb,
preprint,
]{revtex4-2}
\usepackage{color}
\usepackage{ulem}
\usepackage{amsmath}
\usepackage{subfigure}
\usepackage{graphicx}
\usepackage{dcolumn}
\usepackage{bm}

\usepackage[utf8]{inputenc}
\usepackage[T1]{fontenc}
\usepackage{mathptmx}
\usepackage{etoolbox}
\makeatletter
\def\@email#1#2{%
 \endgroup
 \patchcmd{\titleblock@produce}
  {\frontmatter@RRAPformat}
  {\frontmatter@RRAPformat{\produce@RRAP{*#1\href{mailto:#2}{#2}}}\frontmatter@RRAPformat}
  {}{}
}%
\makeatother
\begin{document}


\title{Two-dimensional shaping of Solov'ev equilibrium with vacuum using external coils}
\author{Jiaxing Liu}
\affiliation{
	International Joint Research Laboratory of Magnetic Confinement Fusion and Plasma Physics, State Key Laboratory of Advanced Electromagnetic Engineering and Technology, School of Electrical and Electronic Engineering, Huazhong University of Science and Technology, Wuhan, 430074, China
}
\author{Ping Zhu}%
\email{zhup@hust.edu.cn} 
\affiliation{
	International Joint Research Laboratory of Magnetic Confinement Fusion and Plasma Physics, State Key Laboratory of Advanced Electromagnetic Engineering and Technology, School of Electrical and Electronic Engineering, Huazhong University of Science and Technology, Wuhan, 430074, China
}%
\affiliation{
	Department of Engineering Physics, University of Wisconsin-Madison, Madison, Wisconsin 53706, United States of America
}

\author{Haolong Li}
\affiliation{%
College of Physics and Optoelectronic Engineering, Shenzhen University, Shenzhen 518060, China
}%

\date{\today}

\begin{abstract}
In this work, we demonstrate a method for constructing the Solov'ev equilibrium with any given 2D shape surrounded by a vacuum region using external poloidal field coils.
The computational domain consists of two parts: the plasma region, where the solution is the same as the Solov'ev solution, and the vacuum region, where the magnetic field generated by external coils as well as plasma current is determined using the Green function method through a matching condition near the separatrix. 
However, the method is not limited to the Solov'ev equilibrium in particular. The accuracy, efficiency, and robustness of such a scheme suggest that this method may be applied to the 2D shaping of tokamak plasma with vacuum region using external coils in general.

\end{abstract}

\maketitle



For a tokamak plasma, its stability and transport can be significantly affected by the two-dimensional (2D) shaping of equilibrium poloidal cross-section\cite{Wahlberg11,Aiba_2007,Risk_1981,grauer_rebhan_1984}. Thus it is often desirable to construct the tokamak equilibrium with any prescribed poloidal shape for its plasma-vacuum interface using external coils in both theory and experiments.


2D shaping of tokamak equilibrium has been a subject of extensive studies. For certain pressure and current profiles, several analytical solutions of the Grad-Shafranov (G-S) equation allowing arbitrary plasma shape have been derived, where the shaping parameters, including elongation and triangularity, are determined using the expansion coefficients of the homogeneous solutions \cite{solov1968theory,zheng_analytical_1996,Guazzotto2007AFO,cerfon_one_2010}. 
In simulation and experiments, 2D shaping studies have been often carried out in the design and optimization of equilibrium configuration feedback control schemes, which have been implemented in most tokamak devices (e.g. EAST\cite{Yuan_2013} and DIII-D\cite{DIIIDfeedback}). 
one common such scheme usually involves the measurement and calculation of poloidal flux at specific control points\cite{Hofmann_1990,Albanese2011,optimalchoice}.
For the given plasma current and pressure profiles, any deviation of the poloidal shaping from its target at a certain control surface can be corrected by adjusting the external coil current, and the exact adjustment can be determined by the matching condition at the control points. Such a method has been found successful in practice \cite{DIIIDfeedback,Yuan_2013}.
However, the direct relation between the 2D plasma shape and external coil configurations, which would allow the identification of their overall geometric and physical connections, has yet to be explicitly established. 


This paper presents	 a method for finding the direct and explicit relation between the tokamak shaping and the external poloidal field coil configuration, taking the well-known Solov'ev equilibrium as an example. As one of the analytical solutions of the G-S equation, Solov'ev equilibrium is often used in the benchmark of the numerical G-S equation solver and the analysis of tokamak equilibrium, such as  the study on 2D plasma shaping\cite{zheng_analytical_1996,cerfon_one_2010}.
The conventional Solov'ev solution, however, is derived with toroidal current filling all space, which is in contrast to the tokamak experiment where there is virtually a vacuum out of separatrix. Even though many studies have been done on the relationship between Solov'ev solutions and plasma shape\cite{zheng_analytical_1996,cerfon_one_2010}, most of these studies are based on the conventional Solov'ev solution without vacuum region. 
Xu and Fitzpatrick find an overall equilibrium solution with the Solov'ev solution in plasma along with a vacuum region\cite{2019_xu} using the Green function and multiple expansion method, 
which may fail when the external coils are close to the plasma boundary as in realistic experiments. 
Instead, we directly use the Green function method for both plasma current and external poloidal field coils, which allows us to not only find the matching vacuum solution to a given Solov'ev equilibrium inside the separatrix but also establish the direct and explicit relation between the plasma shape and the external coil configuration. In general, this method can be directly applied to obtain tokamak equilibrium with desired 2D plasma shape in both simulation and experiment. 

The rest of this paper is organized as follows: first, the numerical method is presented for finding the relation between plasma shaping and external poloidal field coil configuration. Then, we demonstrate the usage of the method and evaluate the associated numerical error using an example case of Solov'ev equilibrium. 
Finally, we apply the method to discuss the explicit relations between several typical plasma shaping and their corresponding external poloidal field coil configurations.

\textit{Numerical method:} 
For the MHD equilibrium with toroidal axisymmetry, the poloidal magnetic flux function $\psi$ is governed by the G-S equation 
\begin{equation}\label{equ:norGS}
	R\frac{\partial}{\partial R}\left(\frac{1}{R}\frac{\partial \psi}{\partial R}\right)+\frac{\partial^2 \psi}{\partial Z^2}=-RJ_\phi
\end{equation}
with
\begin{equation}
	RJ_\phi=R^2\frac{dp}{d\psi}-F\frac{dF}{d\psi}
\end{equation}
Here, $\left(R, \phi, Z\right)$ corresponds to the right-hand cylindrical coordinate system. 
The above equation is written in the dimensionless form, where the magnetic field, poloidal magnetic flux, plasma pressure $p$, toroidal current density $J_\phi$, and poloidal current are normalized by the constants $R_0, B_0, R_0^2B_0, B_0^2/\mu_0, B_0/\left(\mu_0R_0\right)$, and $R_0B_0$ respectively, with $R_0, B_0$ being the major radius and magnetic field strength at the magnetic axis.
%

The solutions to the G-S equation consist of the special and homogeneous contributions:
\begin{equation}
	\psi=\psi_0+\sum_{i=1}^{n}c_i\psi_i\qquad\left(i=1, 2, 3..., n\right)
	\label{equ:solution}
\end{equation}
where $\psi_0$ is the special solution specified by the pressure and the current profiles, and $\psi_i$ is a set of homogeneous solutions of the G-S equation (\ref{equ:norGS}) that satisfy certain boundary conditions. The homogeneous solution used here assumes that the overall solution can be written as a Taylor series, starting from a constant and increasing up to and including $n$-th order terms\cite{zheng_analytical_1996,shafronov1968,freidberg_2014,Zakharov1973,Reusch1986}. Next, we can solve the G-S equation by finding the corresponding matching solutions with vacuum in the form of Eq.~(\ref{equ:solution}) from two sub-regions of the computational domain. The first part is the plasma region with closed magnetic field lines, and the second is the vacuum region with open magnetic field lines. 

\textit{$\psi$ in the closed field line region:} 
In this region, we first determined the special solution given by the pressure and current profiles. In the case of Solov'ev equilibrium, the pressure and current profiles are as follows:
\begin{equation}\label{equ:solovev-condition}
	\begin{array}{rl}
		-\frac{dp}{d\psi}&=a\\
		F\frac{dF}{d\psi}& =b
	\end{array}
\end{equation} 


For any given special solution, the 2D plasma shaping is further determined by the homogeneous solution or in terms of its expansion in Eq.~\ref{equ:solution}, a particular set of the coefficients $c_i$. On the other hand, all the desired local and global  geometric features of the plasma shaping can be used to construct the corresponding set of constraints on the flux function or its partial derivatives at locations, such as the X points, the magnetic axis, or the surface, which allows the specification of the homogeneous solution to the extent of an equal number of coefficients $c_i$. This completes the equilibrium solution in the plasma region on the closed field lines inside the separatrix.
\textit{$\psi$ in the open field line region:} The magnetic flux $\psi$ in the open field line region is determined by the plasma current and the external poloidal field coils, which, in general, can be calculated using the Green function method:
\begin{equation}
	\psi=\psi_p+\psi_c
\end{equation}
\begin{equation}\label{equ:psip}
	\psi_{p}(R,Z)=\iint _\Omega G\left(R, Z;R^{\prime},z^{\prime}\right) J_{\phi}\left(R^{\prime}, Z^{\prime}\right) d R^{\prime} d Z^{\prime}
\end{equation}
\begin{equation}\label{equ:coil}
	\psi_{c}(R,Z)=\sum_{i=1}^{N_{c}} G\left(R,Z;R_{i}^{c}, Z_{i}^{c}\right) I_{i}
\end{equation}
where $\Omega$ represents the plasma region inside the separatrix, $J_\phi$ is the toroidal current density, $N_c$ is the total number of the external coils, and $I_i$  is the current in each of the external coils, the Green function  $G\left(R, Z;R^{\prime},z^{\prime}\right)$ is given by \cite{Jardin2010} 
\begin{equation}\label{equ:GR}
	G\left(\mathbf{R} ; \mathbf{R}^{\prime}\right)=\frac{1}{2 \pi} \frac{\sqrt{R R^{\prime}}}{k}\left[\left(2-k^{2}\right) K(k)-2 E(k)\right]
\end{equation}
where $K\left(k\right)$ and $E\left(k\right)$ are the complete elliptic integrals of the first and second kind, and the argument is defined by
\begin{equation}\label{equ:k}
		k^{2}=\frac{4 R R^{\prime}}{\left[\left(R+R^{\prime}\right)^{2}+\left(Z-Z^{\prime}\right)^{2}+\epsilon\right]}
\end{equation}
which can be derived through calculating the vector potential generated by an axisymmetric current ring with unit current\cite{jackson1998classical}. And $\epsilon$ is a small quantity added to avoid the singularity at the control surface caused by the complete elliptic integral of the first kind $K$.

\textit{Matching on last closed field surface:} To guarantee the continuity of poloidal magnetic flux $\psi$ between the open and the closed field line regions, the values of $\psi$ from both regions could be matched on a surface close to the LCFS at chosen matching points $j$
\begin{equation}\label{equ:match}
	\psi_{pj}+\psi_{cj}=\left(1-\eta\right)\psi_X
\end{equation}
where $\psi_X$ is the poloidal magnetic flux at the X point on separatrix, $\psi=\left(1-\eta\right)\psi_X$ is the flux on the matching surface, $\eta$ is a small number chosen to avoid singularities caused by discontinuities in the tangential slopes of the magnetic field lines at the X points\cite{2019_xu}. For the purpose of plasma 2D shaping control, the extent that the exact configuration setting of the external poloidal field coils needs to be specified may be determined by the set of chosen locations where the matching conditions are required to be satisfied.
From the expression of $\psi_{cj}$, Eq. (\ref{equ:coil}-\ref{equ:k}), one can see that the dependence of matching conditions on external coil current is linear, whereas their dependence on external coil location is nonlinear. Thus, it would be more straightforward and practical to solve for the set of external coil current at the fixed coil locations in order to satisfy the set of given matching conditions, which is often the case in reality.

In this way, we obtain the complete equilibrium solution to the G-S equation in Eq.~(\ref{equ:norGS}) where the corresponding configuration of the external poloidal field coils can maintain the prescribed plasma profiles and 2D shape at the separatrix are also determined. 

\textit{An example case and error evaluation:} An up-down symmetric tokamak equilibrium with separatrix, which can be minimally specified with the first four items expansion terms of the homogeneous solutions along with a special solution of the G-S equation (\ref{equ:norGS}), is shown here as an example of the solution method described in the previous section:
\begin{equation}\label{equ:homo}
	\begin{array}{rl}
			\psi_0&=\frac{a}{8}R^4+\frac{b}{2}\left(R^2\ln R-\frac{R^2}{2}\right)\\
			\psi_1&=1, \qquad\qquad\qquad \psi_2=R^2\\
			\psi_3&=Z^2-R^2\ln R, \qquad \psi_4=R^4-4R^2Z^2
		\end{array}
\end{equation}
In this case, the four unknown coefficients $c_1-c_4$ in Eq.~(\ref{equ:solution}) can be determined by the locations of the X points $(R_X,\pm Z_X)$ and the magnetic axis $\left(R_0,0\right)$ together with the assumption that the poloidal flux function  $\psi=0$ at the magnetic axis. This yields the corresponding constraints: 
\begin{equation}\label{equ:symm-cond}
	\begin{array}{rl}
		\frac{\partial\psi}{\partial R}&=0, \quad \frac{\partial\psi}{\partial Z}=0, \qquad\mbox{at X points } (R_X,\pm Z_X)\\
		\hfill\\
		\frac{\partial\psi}{\partial R}&=0, \quad \psi=0, \qquad \mbox{at magnetic axis }\left(R_0,0\right)\\
		\hfill
	\end{array}
\end{equation}
The constraint that the derivative of $\psi$ on $Z$ at the magnetic axis equals zero is naturally satisfied, owing to the symmetry of $\psi$ in $Z$. The constraints in Eq.(\ref{equ:symm-cond}) lead to a set of linear inhomogeneous algebraic equations for the unknown $c_i$ which uniquely specify the 2D shape of the magnetic flux surface in the closed field line region. And the external coil setting designed to maintain the 2D shape of the closed field line region can be determined by the matching conditions in Eq.~(\ref{equ:match}), which would reduce to essentially a set of linear equations for the external coil current $I_i$ at fixed coil locations.

For example, the plasma current and pressure are specified by $a=1.2, b=-1.0$. The shape parameters are chosen as $R_0=1.0, R_X=0.95, Z_X=\pm 0.06$. Corresponding to the up-down symmetry of this particular equilibrium, a number of external coils at fixed locations with up-down symmetry are also prescribed. In this way, we only need to solve half number of the external coil currents. The exact number of the external coils is determined by the number of match points selected. For the Green function integral in Eq.~(\ref{equ:psip}), the two-dimensional trapezoidal rule is adopted on a $100\times100$ uniform rectangular grid.



\begin{figure}[htbp]
	\centering
	\includegraphics[width=0.65\linewidth]{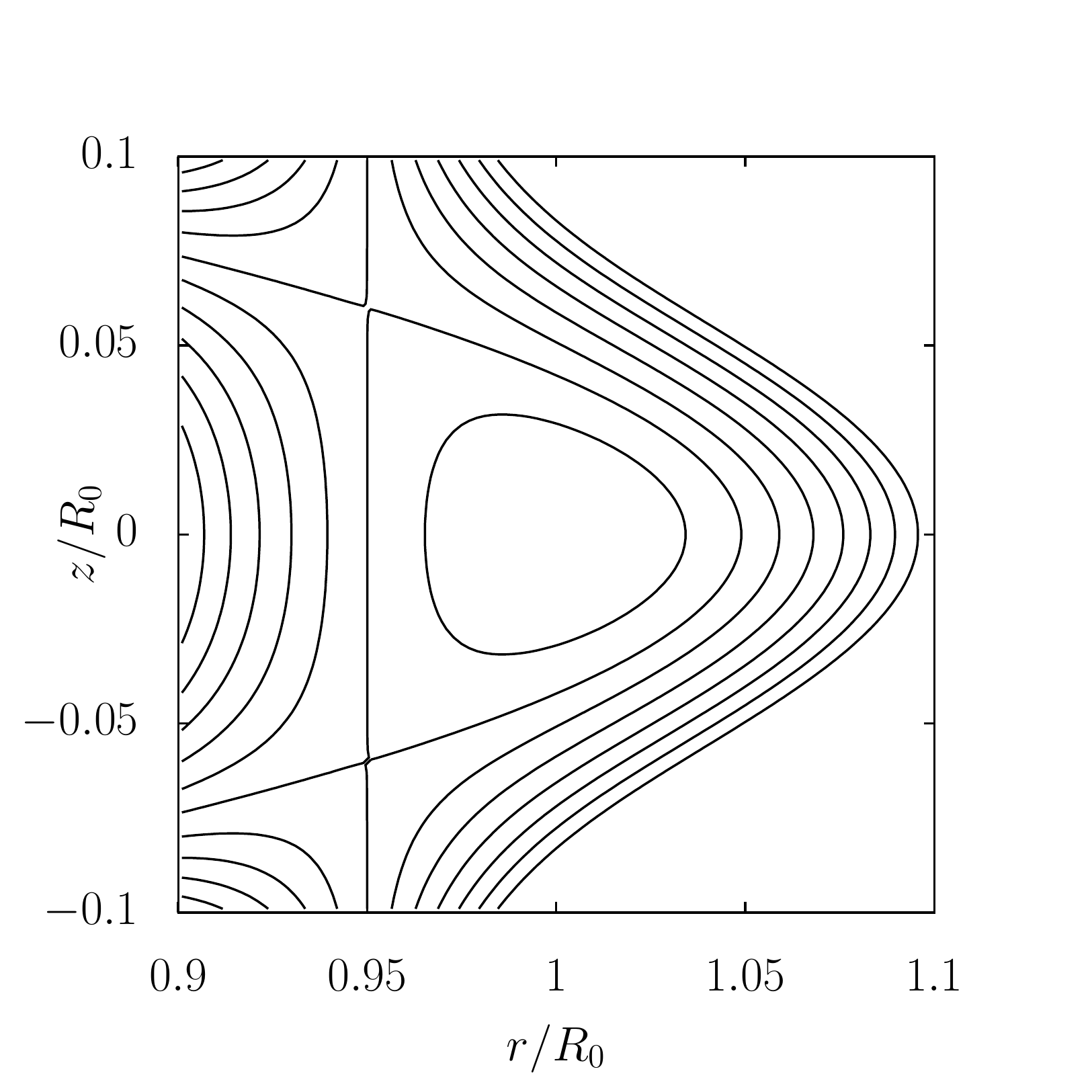}
	\includegraphics[width=0.65\linewidth]{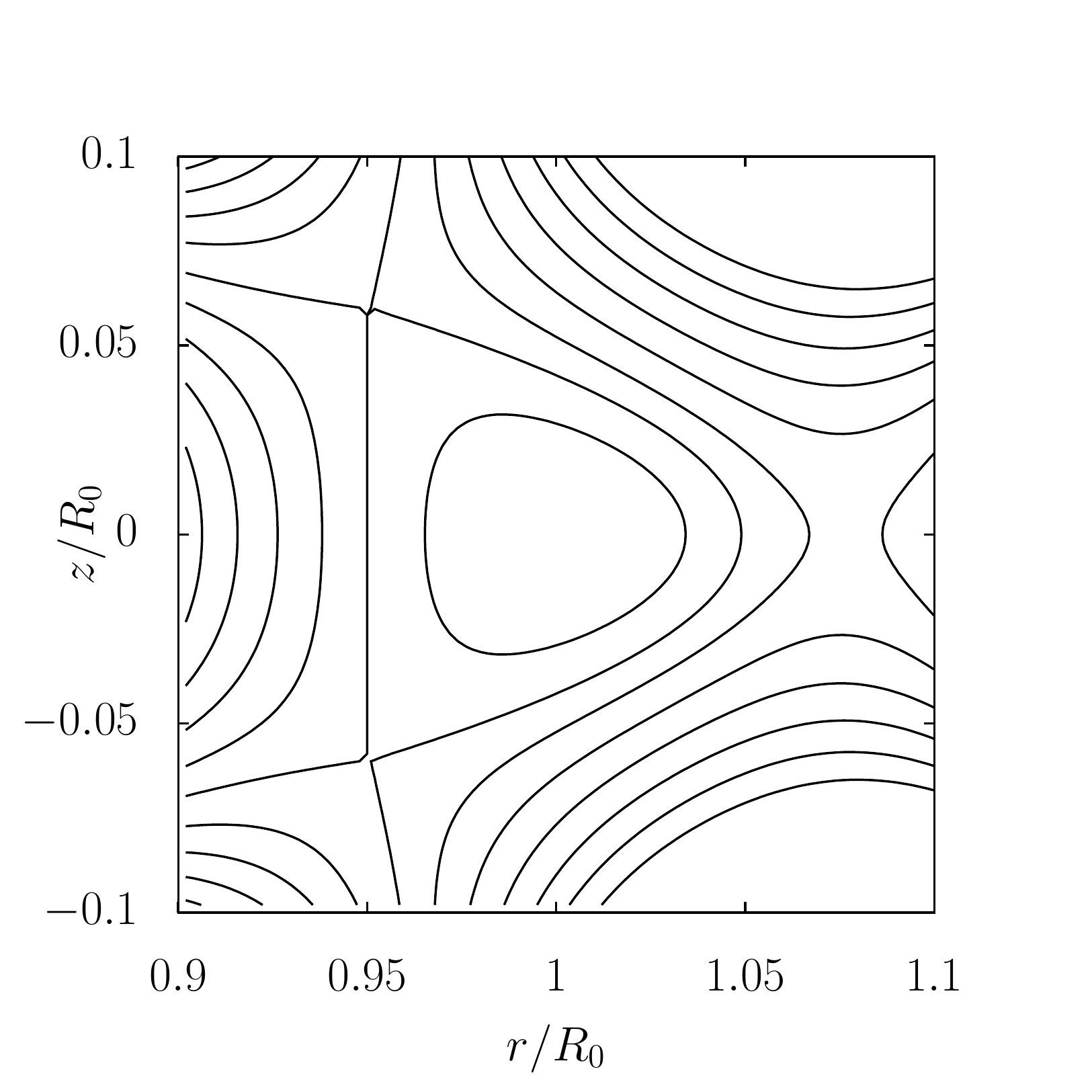}	
	\caption{Contour of $\psi$ in Solov'ev solution with (lower) and without (upper) vacuum region, where $a=1.2$, $b=-1.0$, $R_0=1.0$, $R_X=0.95$, $Z_X=\pm 0.06$ as in Eq.~(\ref{equ:homo}) and Eq.~(\ref{equ:symm-cond})}
	\label{fig:sol-sym}
\end{figure}
\begin{figure}[htbp]
	\centering
	\includegraphics[width=0.65\linewidth]{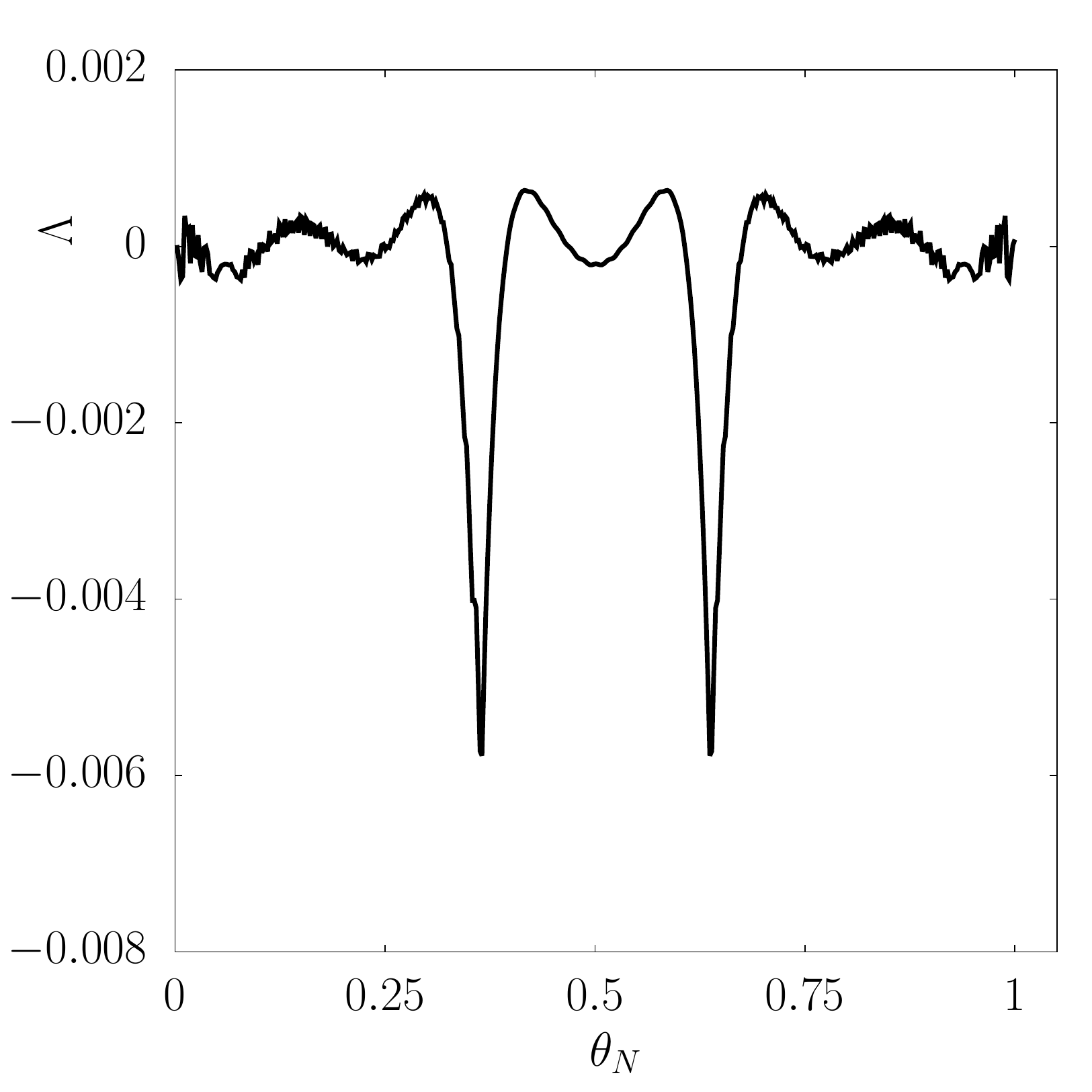}
	\includegraphics[width=0.65\linewidth]{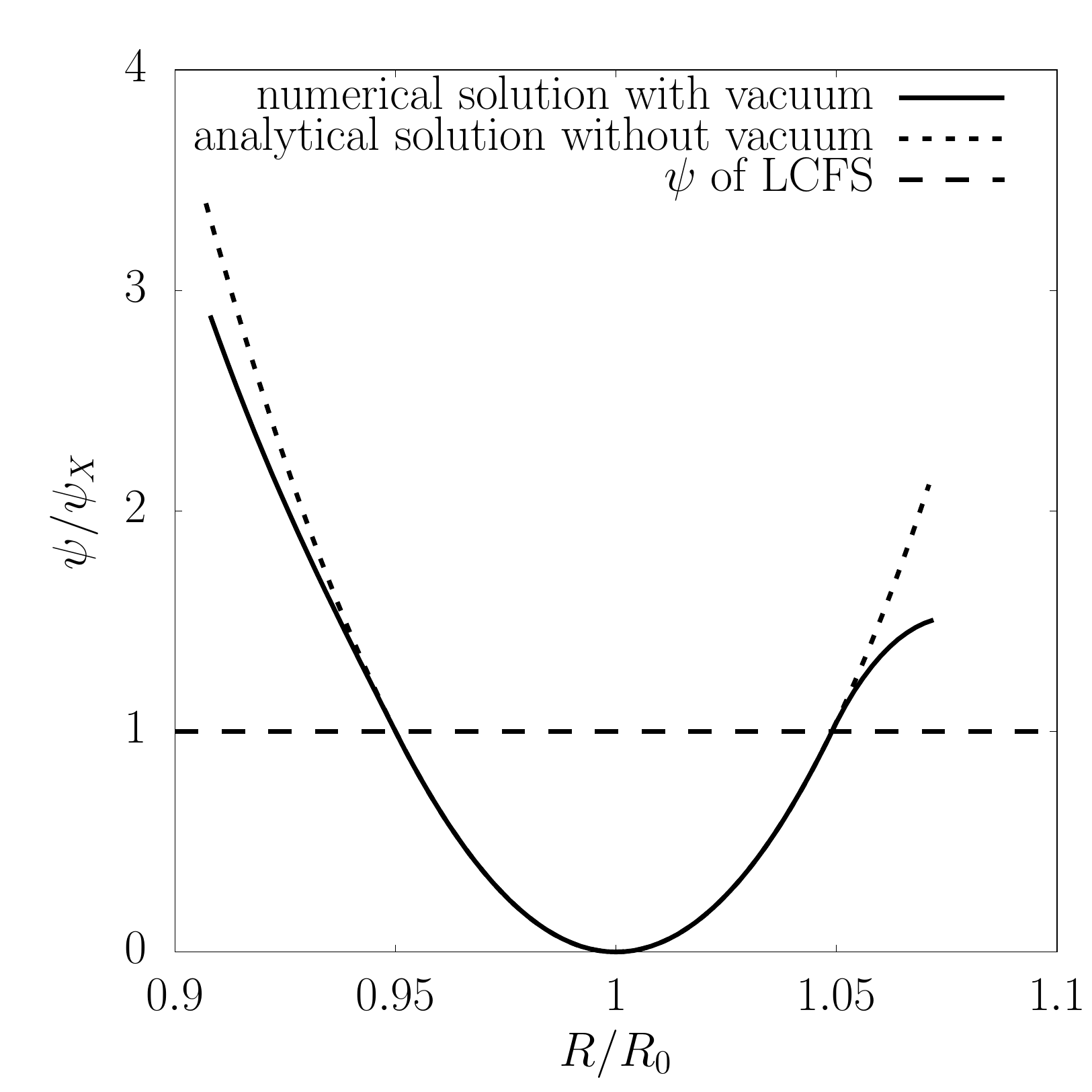}
	\caption{The relative difference  in $\psi$ between the Solov'ev equilibrium solutions with and without vacuum on the middle plane (lower) and on the LCFS (upper) as a function of the normalized poloidal angle $\theta_N$.}
	\label{fig:diff-sym}
\end{figure}
\begin{figure}[htbp]
	\centering
	\includegraphics[width=0.65\linewidth]{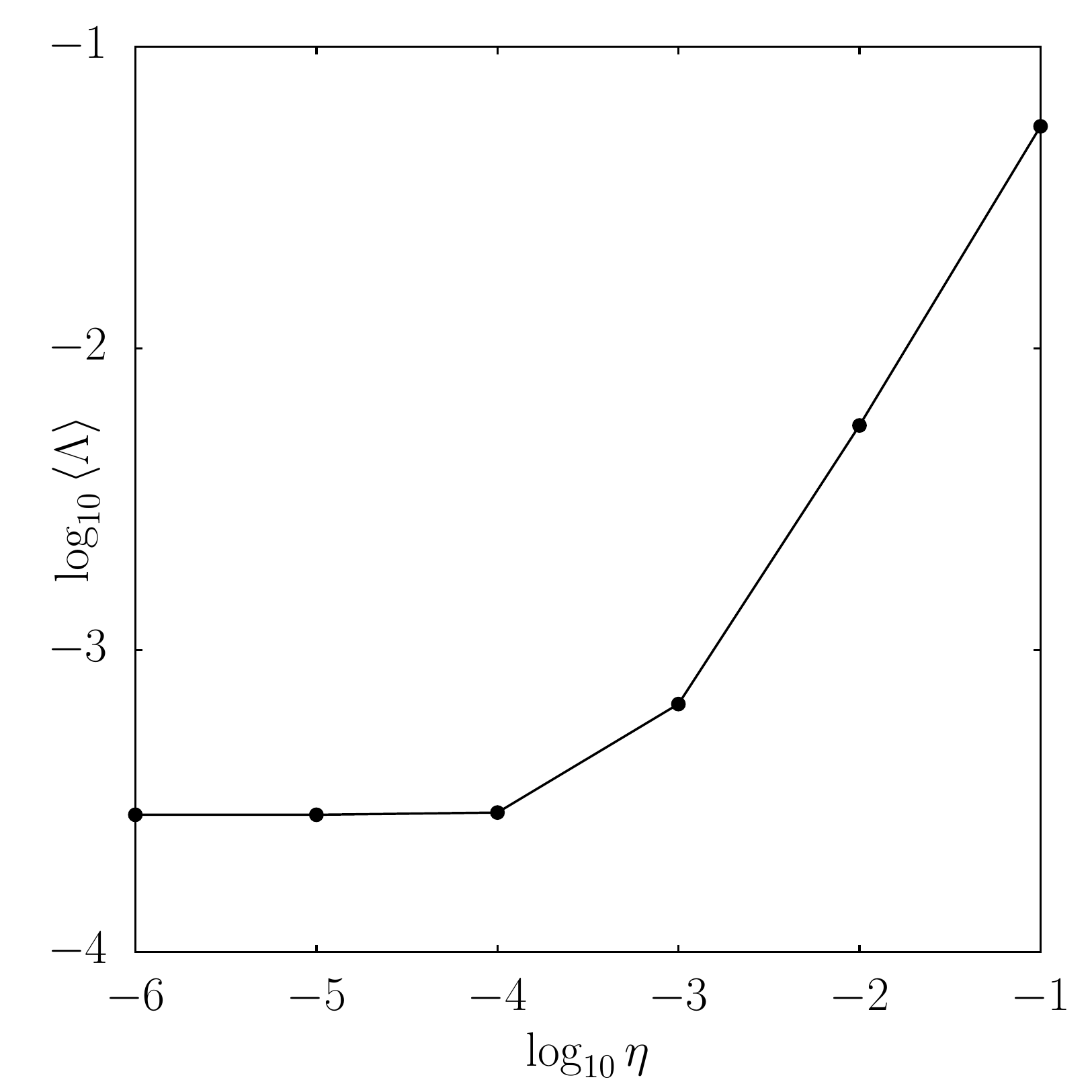}
	\caption{The averaged relative error $\left<\Lambda\right>$ between the Solov'ev equilibrium with and without vacuum on the LCFS as a function of $\log_{10}\eta$ for a fixed grid size.}
	\label{fig:ralative-error}
\end{figure}

Using the method outlined in previous sections, we obtain the Solov'ev equilibrium with a vacuum region outside the separatrix (Fig.~\ref{fig:sol-sym}) and the corresponding external coil configuration. In comparison to the conventional Solov'ev  equilibrium without a vacuum region, the contours of $\psi$ are the same in the closed field line region, and the shape of the separatrix is nearly identical subject to the numerical error. 
The relative error of the vacuum solution on the separatrix can be defined by   $\Lambda=\frac{\psi_X^{vac}(\theta)-\psi_X}{\psi_X}$, where $\psi_X$ is the Solov'ev solution on the separatrix and $\psi_X^{vac}(\theta)$ is the numerical vacuum solution along the separatrix and dependent on the poloidal angle $\theta$ at the given $\psi_X$ on the separatrix.
As shown in the upper figure of Fig.~\ref{fig:diff-sym}, the maximum difference is at the X points, which is caused by the discontinuity of the tangential slope of the magnetic field line at the X points. Nonetheless, as seen in the lower figure of Fig.~\ref{fig:diff-sym}, $\psi$ is continuous at separatrix. Outside the separatrix, the contour of $\psi$ differs from the Solov'ev solution, as also indicated by the different external angles of field lines at the X points. 
The error $\Lambda$ depends on the matching parameter $\eta$ and the grid size. Keeping the grid size fixed, the relative error increases with $\eta$, as shown in Fig.~\ref{fig:ralative-error}. For several fixed given $\eta$, the numerical convergence in terms of $\Lambda$ with grid size is also confirmed.  



\textit{2D shaping by external coils:} In the above method, only a set of linear equations for the coil currents need to be solved for a set of prescribed coil locations, which suggests that the method may be used as an efficient 2D shaping control scheme using external coils.  
To demonstrate this, we solve for example the external coil settings for several Solov'ev equilibria with vacuum regions and the designed features of 2D shape, such as varying degrees of up-down asymmetry, elongation, and triangularity. 



\textit{Up-down asymmetry:} 
For an up-down asymmetric equilibrium, the positions of the magnetic axis and two different X points can be chosen to determine the plasma shape. As a result of the choice, we  will need to select at least three additional expansions terms of the homogeneous solutions of Eq.~(\ref{equ:norGS}). 
\begin{equation}
	\begin{array}{rl}
		\psi_5& =Z,\qquad\qquad\qquad \psi_6=R^2Z,\\
		\psi_7&=Z^3-3ZR^2\ln R
	\end{array}
\end{equation}
Now, seven constraints are required to solve the seven unknown coefficients $c_i$. 
\begin{equation}\label{equ:asymm-cond}
	\begin{array}{rl}
		\frac{\partial\psi}{\partial R}&=0, \quad \frac{\partial\psi}{\partial Z}=0, \qquad\mbox{two X points}~(R_{x1},Z_{x1}),~(R_{x2},Z_{x2})\\
		\hfill\\
		\frac{\partial\psi}{\partial R}&=0, \quad \frac{\partial\psi}{\partial Z}=0,\quad \mbox{magnetic axis}~(R_0,0)\\
		\hfill\\
		\psi&=0,\qquad \mbox{magnetic axis}~(R_0,0)
	\end{array}
\end{equation}

\begin{figure}[htbp]
	\centering
	\subfigure{
	\includegraphics[width=0.65\linewidth]{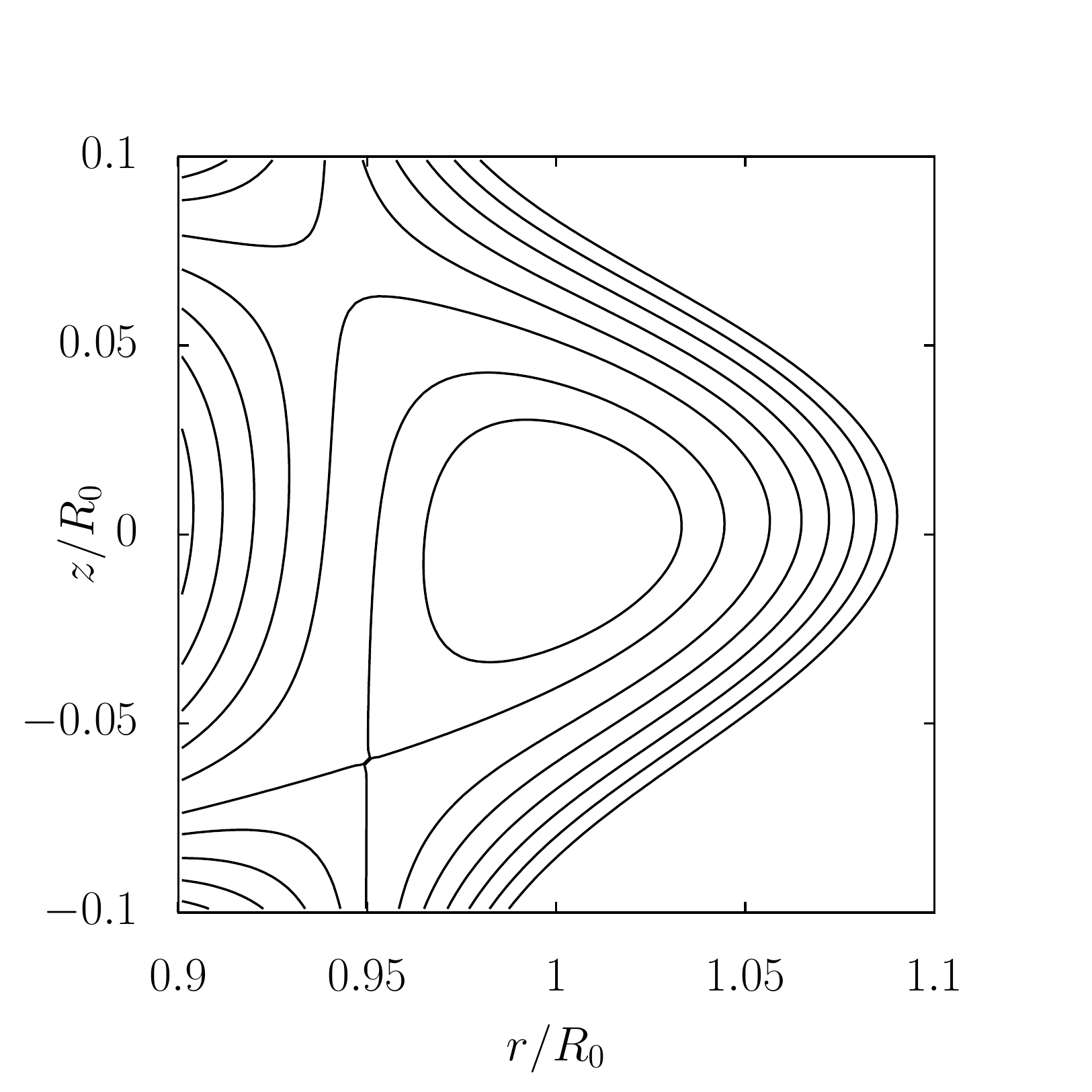}
	}
	\subfigure{
	\includegraphics[width=0.65\linewidth]{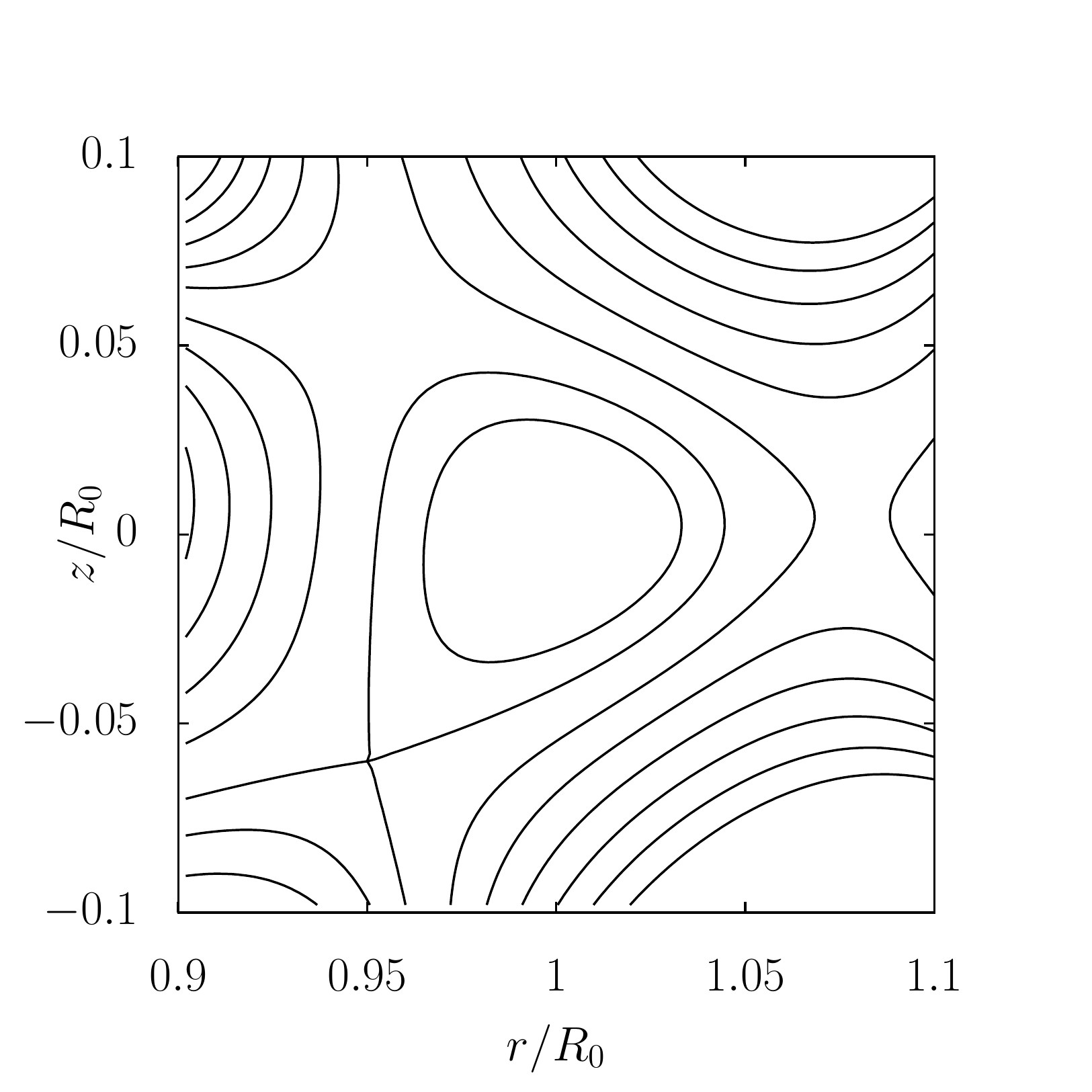}
	}
	\caption{
		Contour of $\psi$ in Solov'ev solution with (lower) and without (upper) vacuum region, where $R_{x}^{\mbox{upper}}=0.94, Z_{x}^{\mbox{upper}}=0.07, R_{x}^{\mbox{lower}}=0.95, Z_{x}^{\mbox{lower}}=-0.06, R_0=1.0,Z_0=0$ as in Eq.~(\ref{equ:asymm-cond}) and there are also fourteen poloidal field coils.}
	\label{fig:assy-sol}
\end{figure}

The constraint on the derivative of $\psi$ on Z at the magnetic axis is required now because it is not satisfied automatically due to the loss of symmetry in Z. Similarly, the currents of all coils, instead of half, must be calculated.
For the same external coil locations, and the plasma current and pressure profiles in the up-down symmetric Solov'ev equilibrium specified in Eq.~(\ref{equ:homo}), the conventional Solov'ev equilibrium and the Solov'ev equilibrium with a vacuum region with the same plasma shape are shown in Fig.~\ref{fig:assy-sol} respectively.

\begin{figure}
	\centering
	\includegraphics[width=0.65\linewidth]{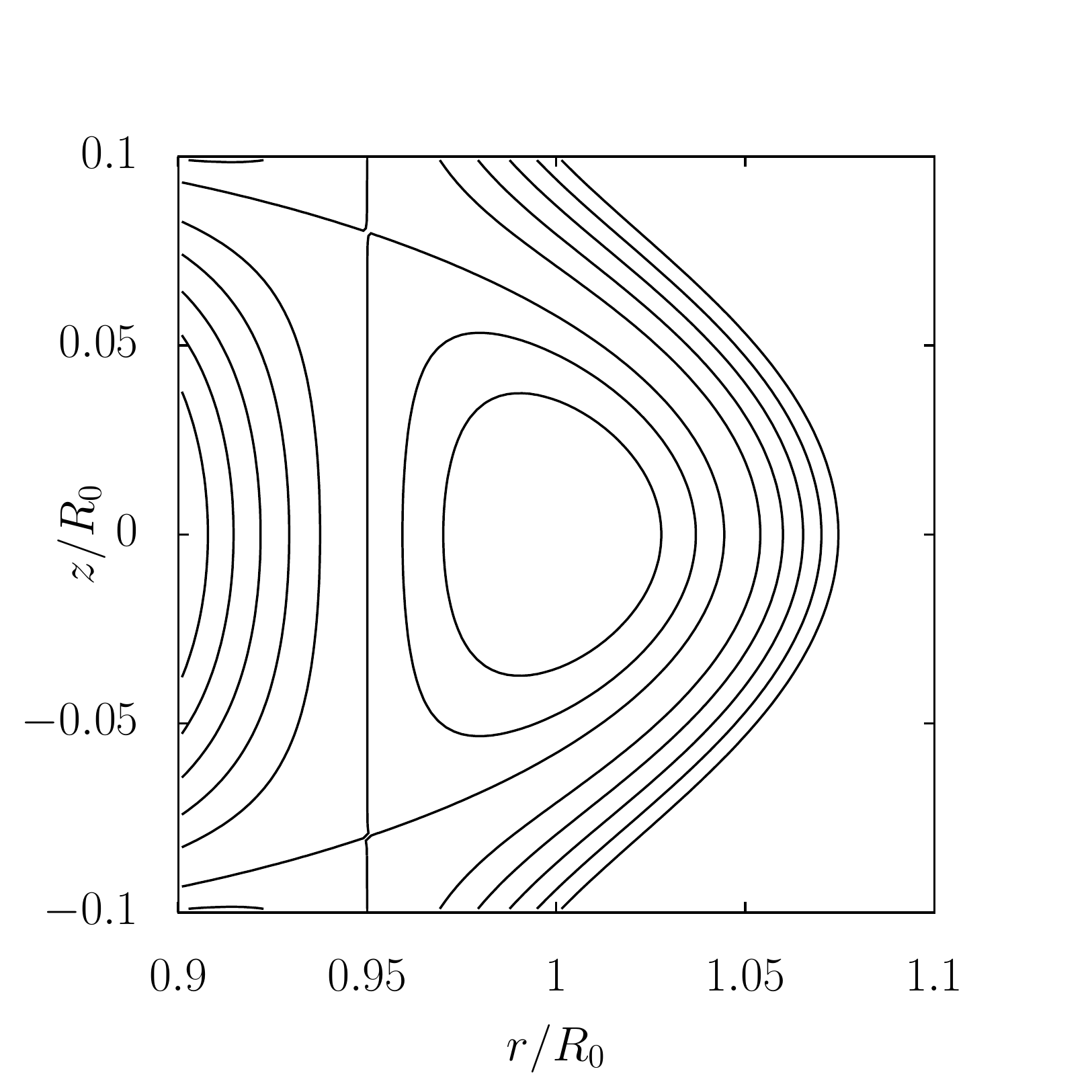}
	\includegraphics[width=0.65\linewidth]{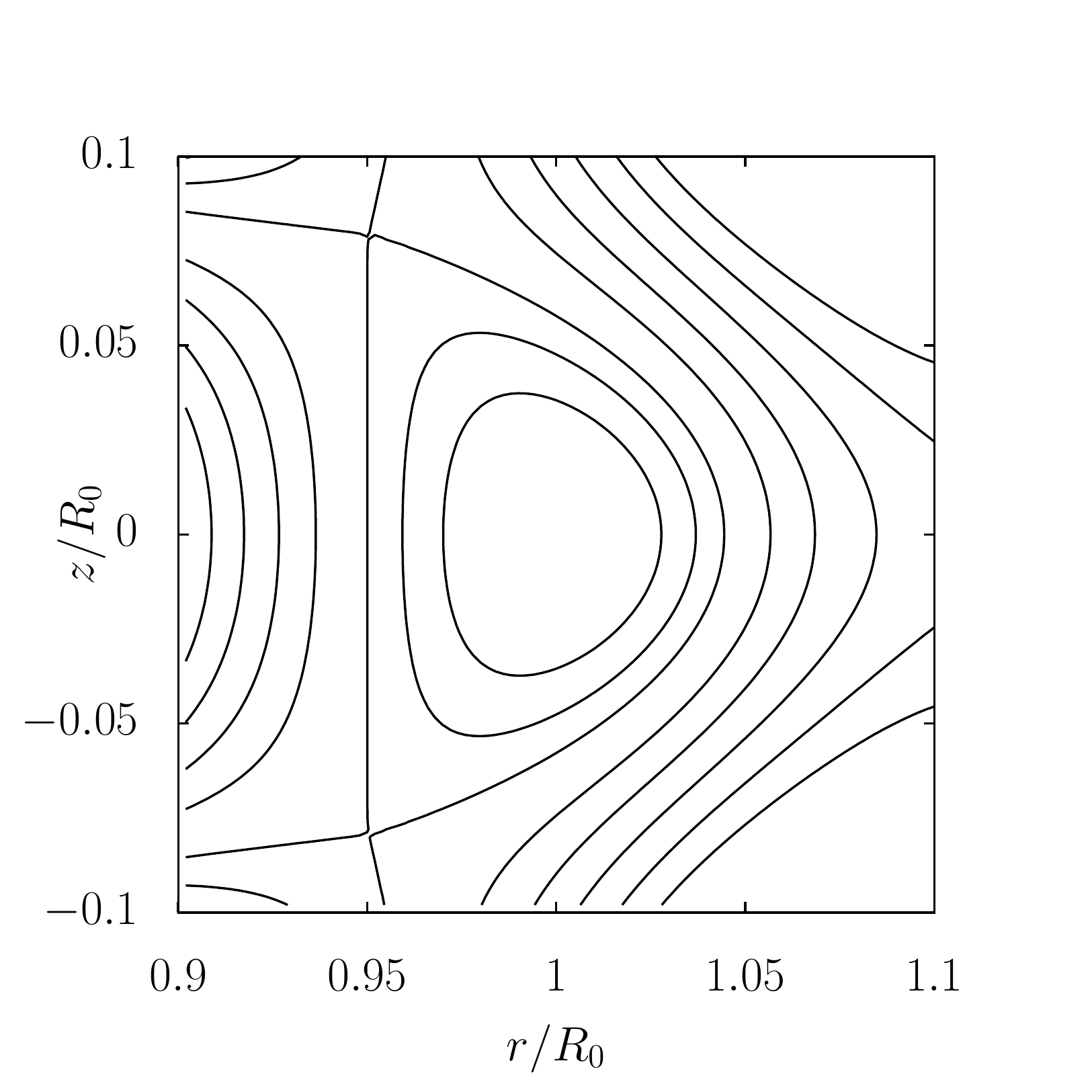}
	\caption{Contour of $\psi$ in Solov'ev solution with (lower) and without (upper) vacuum region, where $a=1.2$, $b=-1.0$ as in Eq.~(\ref{equ:homo}) for a larger elongation with $R_X=0.95, Z_X=\pm 0.08, R_0=1.0$ as in Eq.~(\ref{equ:symm-cond}).}
	\label{fig:sol-largeE}
\end{figure}

\textit{Elongation:} For an up-down symmetry equilibrium, the position of the X points can be used to change the elongation. 
Using the same constraints as in Eq.~(\ref{equ:symm-cond}) and different positions of X points, an equilibrium with different elongation is obtained which is shown in Fig.~\ref{fig:sol-largeE} along with the classical Solov'ev equilibrium without vacuum for comparison.

\begin{figure}[htbp]
	\centering	
	\includegraphics[width=0.65\linewidth]{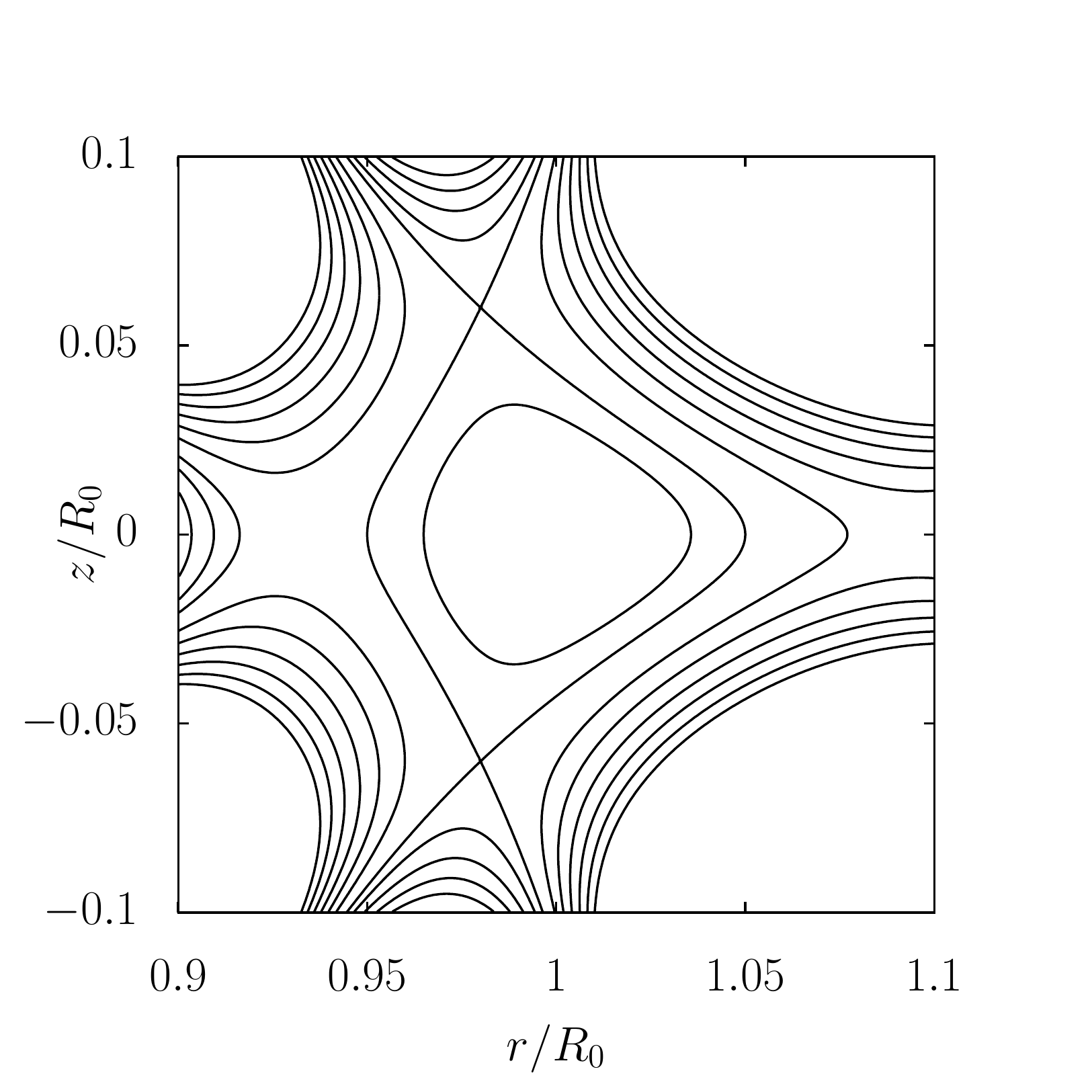}
	\includegraphics[width=0.65\linewidth]{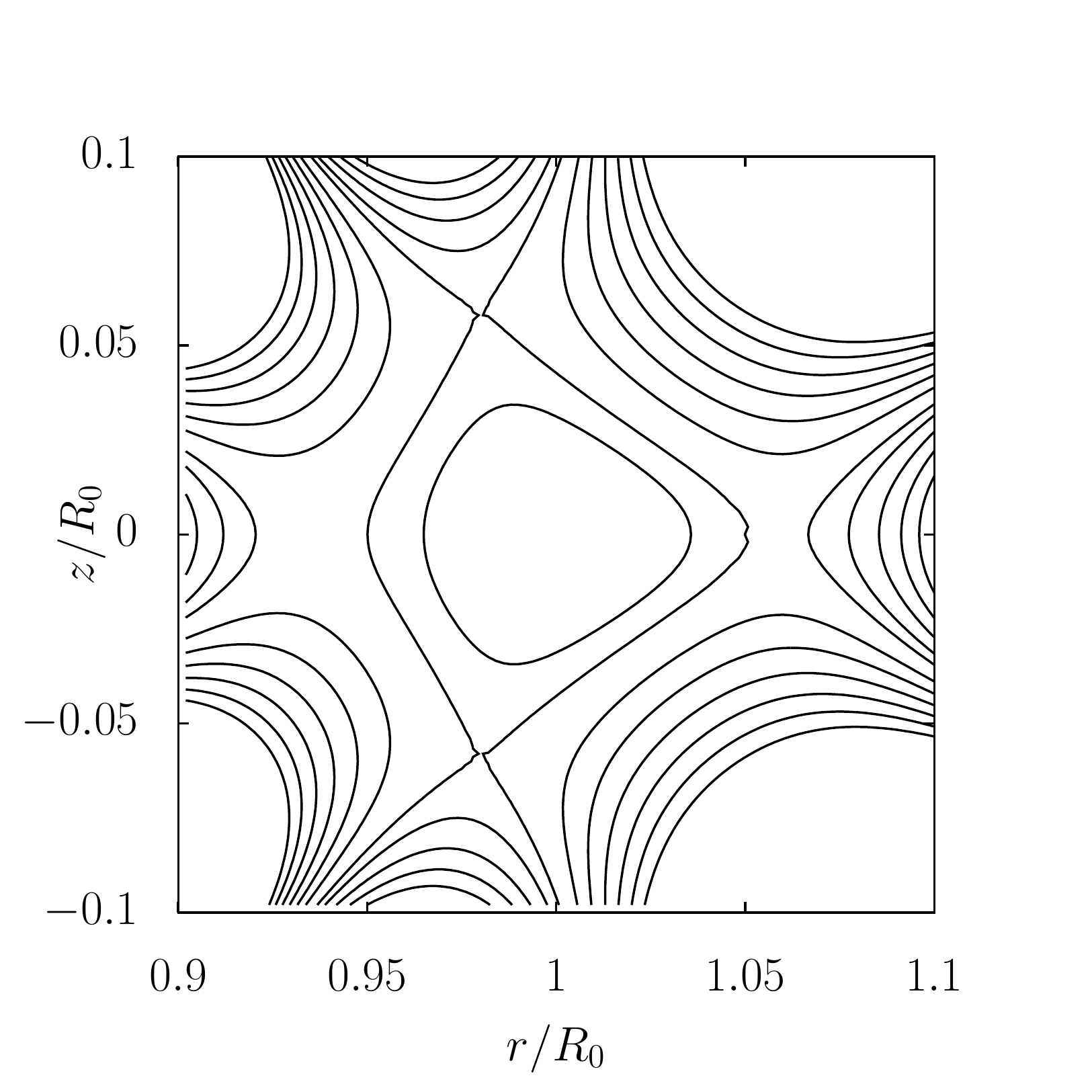}
	\caption{
		Contour of $\psi$ in Solov'ev solution with (lower) and without (upper) vacuum region, where $a=1.2$, $b=-1.0$ as in Eq.~(\ref{equ:homo}), $R_X=0.98, Z_X=\pm 0.06, R_0=1.0 $ as in Eq.~(\ref{equ:symm-cond}), and $R_{inner}=0.95, R_{outer}=1.05$ as in Eq.~(\ref{equ:homo}) and Eq.~(\ref{equ:symm-cond-tri}).} 
	\label{fig:sol-tri}
\end{figure}

\textit{Triangularity:} It can be shown that if only the first four homogeneous solutions of the G-S equation, stated in Eq.~(\ref{equ:homo}), are chosen, the triangularity of Solov'ev equilibrium with X points is equal to 1.
To vary the triangularity of equilibrium, another two expansions terms of the homogeneous solutions of the G-S equation are added
\begin{equation}
\begin{array}{rl}\label{equ:home-tri}
		\psi_5&=2Z^4-9Z^2R^2-12Z^2R^2\ln R+3R^4\ln R \\
		\psi_6&=R^6-12R^4Z^2+8R^2Z^4
\end{array}
\end{equation}
Two corresponding constraints are required to determine these two unknown coefficients, such as the requirement that  $\psi$ at the inner and outer points of LCFS be the same as $\psi$ at the X point
\begin{equation}\label{equ:symm-cond-tri}
	\begin{array}{rl}
		\psi_{inner}&=\psi_X, \qquad \mbox{inner points } (R_{\text{inner}},0)\\
		\psi_{outer}&=\psi_X, \qquad \mbox{outer points } (Z_{\text{inner}},0)
	\end{array}
\end{equation}
The Solov'ev equilibrium thus solved is shown in Fig.~\ref{fig:sol-tri} along with the conventional Solov'ev equilibrium with the same plasma current, pressure profiles, and plasma shape.


In summary, a general method for constructing an equilibrium with any specified 2D shape surrounded by a vacuum region using external poloidal field coils is presented, which allows the determination of the direct relation between the plasma shaping and the corresponding external poloidal field coil setting. And this method is demonstrated to be effective for specifying the elongation, triangularity, and up-down symmetry of Solov'ev equilibrium with a vacuum region by adjusting the current of a set of external coils at fixed locations, for example. However, this method is not limited to the Solov'ev equilibrium. We plan on implementing the method into 2D equilibrium solvers, such as NIMEQ\cite{howell_solving_2014,LINIMEQ} and CHEASE\cite{lutjens_chease_1996}, to obtain a desirable equilibrium with designed plasma shape using external coils and extending this method to the calculation of external coil setting for more general and realistic tokamak equilibrium with any prescribed 2D shape, which may be eventually developed into a practical 2D shaping control scheme.
\begin{acknowledgments}
The author Jiaxing Liu thanks Mr. Rui Han (University of Science and Technology of China) for helpful discussions. This work was supported by the National Key Research and Development Program of China Grant No. 2019YFE03050004, the Fundamental Research Funds for the Central Universities at Huazhong University of Science and Technology Grant No. 2019kfyXJJS193, the National Natural Science Foundation of China Grant No. 51821005, and the U.S. Department of Energy Grant Nos. DE-FG02-86ER53218 and DE-SC0018001. The computing work in this paper is supported by the Public Service Platform of High Performance Computing by Network and Computing Center of HUST.
\end{acknowledgments}

\section*{Data Availability Statement}
The data that support the findings of this study are available from the corresponding author upon reasonable request.

%
%
%
%
%
%
%
%
%

\nocite{*}
\bibliography{references}

\newpage

%

\newpage


\newpage

\newpage


\newpage


\newpage
\end{document}